# Yellow stimulated emission from $Dy^{3+}$-doped silica glass microspheres

ABHISHEK SURESHKUMAR[1], JONATHAN DEMAIMAY[2], GEORGES PERIN[1], SHAHAZ HAMEED[1], MOHAMMED GUENDOUZ[1], HÉLÈNE OLLIVIER[1], YANNICK DUMEIGE[1], PAVEL LOIKO[2], GURVAN BRASSE[2], ALAIN BRAUD[2], PATRICE CAMY[2], AND STÉPHANE TREBAOL[1,*]

[1]*Université de Rennes, CNRS, Institut FOTON—UMR 6082, 6 rue de Kerampont, 22300 Lannion, France*
[2]*Centre de Recherche sur les Ions, les Matériaux et la Photonique (CIMAP), UMR 6252 CEA-CNRS-ENSICAEN, Université de Caen Normandie, 6 Boulevard Maréchal Juin, 14050 Caen, France*
*\*Corresponding author: stephane.trebaol@univ-rennes.fr*

**Abstract:** $Dy^{3+}$-doped silica glass whispering gallery mode microspheres are fabricated by fiber fusion splicing. They present an almost ideal spherical morphology with a radius ranging from 60 to 67 μm as determined by confocal laser microscopy. The host composition of the microsphere is close to that of the fiber core. The dopant $Dy^{3+}$ ions are uniformly distributed across the microsphere as evidenced by μ-luminescence studies and present a luminescence lifetime of the $^4F_{9/2}$ state of 514.7 ± 1.1 μs indicating a weak luminescence quenching. The $Dy^{3+}$-doped glass microspheres were excited via evanescent field coupling using a half-tapered fiber and a blue 450-nm GaN laser diode (direct pumping scheme). The yellow fluorescence of $Dy^{3+}$ ions (the $^4F_{9/2} \to {^6H_{13/2}}$ transition) is filtered by the whispering gallery modes (free spectral range: 0.5 nm for 67-μm radius microsphere). The onset of stimulated emission is further observed highlighting the potential of such microresonators for narrow-linewidth light sources directly emitting visible light.

## 1. Introduction

Coherent light sources directly emitting in the visible spectral range have emerged as an indispensable element in contemporary science and technology. Their ability to emit light with a well-defined phase, amplitude, and polarization has enabled a wide range of applications from high-speed optical communications [1] to quantum information processing [2]. In quantum optics, these sources are used for manipulating and controlling individual atoms and ions [3,4]. Moreover, they have applications extended to biophotonics, imaging, and sensing [5], where compact sources with narrow linewidth [6] are necessary.

Optical microresonators confining light by total internal reflection (TIR) at their dielectric sidewalls are versatile photonic microstructures featuring small size, wavelength selectivity, and resonance field enhancement. A whispering gallery mode (WGM) microresonator is a small, circular structure that confines light, allowing it to circulate along its curved boundary by TIR. WGM resonance behavior could be achieved in dielectric microspheres [7] or microdisks [8]. Such devices with a high-quality factor (Q-factor) and small mode volume (high finesse) offer high energy density inside the cavity leading to a strong light-matter interaction [9–11]. On the other hand, rare earth ions often exhibit complex energy-level schemes determined by their particular electronic configuration $[Xe]4f^n$, allowing them to emit light at wavelengths ranging from ultraviolet (UV) to mid-infrared (IR) [12]. Incorporating such ions into high Q-factor microresonators can enable the development of coherent sources operating at a low threshold and narrow linewidth [13].

In most of the previous works, microlasers operating in the near and mid-IR windows have been reported [14–18]. There also exist studies on microlasers operating in the visible spectrum, excited via the frequency up-conversion mechanism. Fujiwara *et al.* reported on upconversion lasing at 480 nm from a $Tm^{3+}$-doped $ZrF_4$ glass microsphere [19]. Von Klitzing *et al.* achieved low-threshold upconversion lasing at 550 nm from $Er^{3+}$ ions in a fluoride glass microcavity [20]. Wang *et al.* demonstrated laser operation at 545 nm using an $Er^{3+}/Yb^{3+}$-codoped fluorosilicate glass microsphere [21]. The development of blue Gallium Nitride (GaN) laser sources recently enabled the direct pumping of rare earth elements in the visible spectrum with great efficiency [22]. This pumping scheme has already been demonstrated for visible fiber lasers [23]. In the present work, we propose to study the direct pumping of rare earth ions incorporated in a microsphere for visible light emission.

Most of reported microsphere laser studies employed active microresonators fabricated from rare-earth doped bulk glasses using a plasma torch [24] or produced from passive resonators by surface sol-gel deposition [25]. These fabrication techniques are complex, and annealing at high temperatures is often required to ensure the stability of the cavities. A simple technique to fabricate microresonators is to make use of rare-earth doped fibers [14,26]. By melting the fiber tip, *e.g.*, using a CO2 laser [14] or a fiber fusion slicer [27], a rare earth-doped microresonator can be readily obtained. So far, only a few works have employed this technique and there is no clear discussion on the effect of the residual cladding on the possible ion concentration homogeneity in the microresonator.

In the present work, we have used a dysprosium ($Dy^{3+}$) doped aluminosilicate glass fiber to fabricate WGM microresonators using a fusion splicer. In this fabrication process, the two electrodes of the fiber fusion splicer act as a heat source, and by appropriately adjusting their driving current for a specific time duration, the glass transition temperature can be reached. Once the tip of the fiber is heated using the electrical arc generated from the electrodes, the surface tension forces the fiber to form a spherical microresonator.

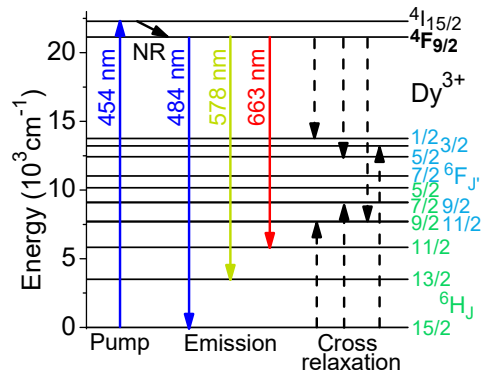

Fig. 1. Energy-level scheme of the $Dy^{3+}$ ion showing pump and luminescence transitions, NR – non-radiative relaxation.

We focus mainly on studying the fabrication technique of microresonators and their structural parameters for lasing in the visible (yellow) spectral range by direct pumping of $Dy^{3+}$ ions in the blue, cf. Fig. 1, benefiting from the well-developed technology of GaN diode lasers. Laser operation of $Dy^{3+}$ silica and fluoride glass fiber lasers directly pumped at 450 nm has been demonstrated recently [28–30]. Several rare-earth ions including $Dy^{3+}$ exhibit absorption bands in the blue spectral range [22], which enables their direct pumping offering an important advantage over the upconversion pumping by IR lasers described above.

## 2. Fabrication of Microspheres

$Dy^{3+}$-doped silica glass WGM microspheres are fabricated by fiber fusion splicing using a commercial single-clad aluminosilicate glass fiber (*Exail*, France). The core/cladding diameters are 25/125 µm, the numerical aperture (N.A.) is 0.18, and the $Dy^{3+}$ doping level in the core is 3000 ppm (0.3 mol%). This relatively low $Dy^{3+}$ doping concentration is selected to prevent excessive self-quenching of the upper laser level lifetime by cross-relaxation.

The fiber is first characterized by µ-luminescence and µ-Raman spectroscopy using a Renishaw in Via Reflex confocal laser microscope equipped with an $Ar^+$ ion laser ($\lambda$ = 457 nm, 514 nm) and a ×50 Leica microscope objective. The fiber end-facet is cleaved, and three different areas are considered: i) the core, ii) the core-cladding interface, and iii) the cladding.

The µ-Raman spectra, Fig. 2(a), exhibit a broad band at 447 cm$^{-1}$ which is associated with the symmetrical Si-O-Si stretching, two well-defined peaks at 489 and 604 cm$^{-1}$ commonly called defect lines $D_1$ and $D_2$ that correspond to the symmetric oxygen breathing of the 4 and 3-membered siloxane rings of the $SiO_4$ tetrahedra, respectively, and at high frequencies, two weak bands appear at 819 cm$^{-1}$ and 1027 cm$^{-1}$, being characteristic of the Si-O asymmetric stretching of tetrahedral units [31]. The µ-Raman spectra are very close for the three studied areas.

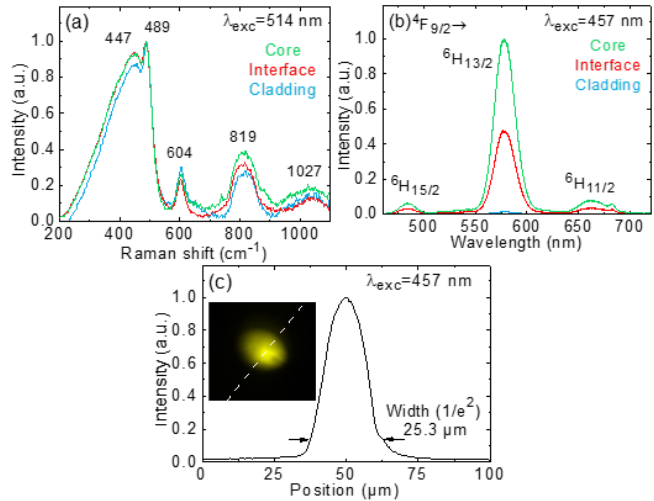

Fig. 2. Spectroscopy of the Dy3+-doped aluminosilicate glass fiber: (a) µ-Raman spectra, λexc = 514 nm; (b) µ-luminescence spectra, λexc = 457 nm (measured from the fiber core, cladding and their interface); (c) profile of intensity of yellow luminescence at 578 nm across the core area, inset – 2D map.

The µ-luminescence spectrum is presented in Fig. 2(b). It reveals three emission bands in the visible, at 484 nm (cyan, the $^4F_{9/2} \rightarrow {}^6H_{15/2}$ transition), 578 nm (yellow, $^4F_{9/2} \rightarrow {}^6H_{13/2}$) and 661 nm (red, $^4F_{9/2} \rightarrow {}^6H_{11/2}$), with the yellow emission being the most intense one and corresponding to an emission bandwidth (full width at half maximum (FWHM) of 23.4 nm. The results of the µ-luminescence mapping monitoring the yellow emission are shown in Fig. 2(c), revealing both the 2D map and the intensity profile across the fiber core. Given the small $Dy^{3+}$ doping level, this mapping could be used to evaluate the distribution of dopant ions in the fiber. The analysis reveals a nearly symmetric $Dy^{3+}$ distribution in the core following a Gaussian-like spatial profile with a width of 25.3 µm (at the $1/e^2$ level).

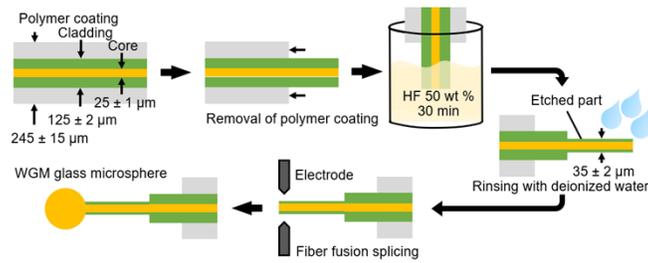

Fig. 3. Schematic of fabricating WGM Dy3+-doped glass microspheres by fiber fusion splicing.

Figure 3 depicts the principal steps of fabrication of WGM $Dy^{3+}$-doped glass microspheres. At the initial step, a 15 cm long fiber is used, and its protective polymer coating is manually removed, followed by selective removal of the cladding layer by immersing the fiber in a 50 % hydrofluoric (HF) acid solution for 30 min. After the etching process, the fiber is rinsed with deionized water to remove the residual HF acid. By examining the fiber region exposed to the acid, we confirm the high surface quality of the processed fiber having a diameter of 35 μm. The microresonator is then fabricated by fiber fusion splicing [27]. The glass transition temperature is achieved by adjusting the driving current to the electrodes. The fiber tip is heated using the electrical arc generated from the electrodes, and the surface tension forces the fiber to form a spherical microresonator.

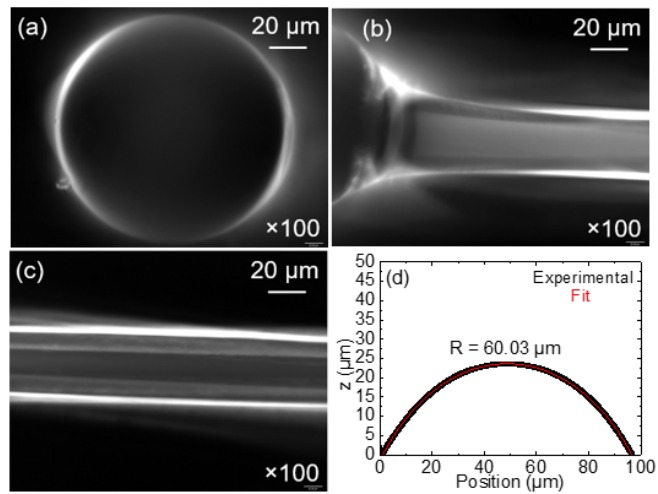

Fig. 4. Confocal microscope study of the $Dy^{3+}$-doped glass microsphere: (a) top view, focusing on the microsphere equatorial plane; (b,c) side view: (b) microsphere/fiber junction; (c) clean treated fiber; (d) surface topography, *black* – measured data, *red* – parabolic fit yielding the radius *r*.

The morphology of the glass microspheres is revealed using a confocal microscope (*Sensofar*, S-neox) using a UV lamp (λ = 365 nm) as illumination source and a ×100 objective, Fig. 4. By setting the imaging plane to the equatorial one to observe the cross-section of the microsphere (in the top view configuration), Fig. 4(a), a perfect spherical morphology and a high homogeneity of the inner area is confirmed. Figure 4(b) reveals the junction between the microsphere and the processed fiber showing just a few ripples. Figure 4(c) shows the processed area of the fiber clearly revealing the difference between the core and the residual cladding. From these observations, we conclude that the material in the microsphere is thoroughly mixed at the stage of fiber fusion as no traces of residual core/cladding interface could be seen. By operating the microscope in the interferometric mode, the surface topography is reconstructed, Fig. 4(d), and well fitted with a parabolic function yielding a radius *r* of 60.03 ± 0.02 µm. No macroscopic defects are revealed at the surface, such as cracks, inclusions, or secondary phases.

To determine the microresonator Q-factor, the cavity ringdown measurement technique is setup up [27]. This technique allows for accurate determination of the intrinsic and extrinsic losses, and the coupling regime of microresonators. The cavity ringdown measurement of the $Dy^{3+}$-doped silica glass WGM microsphere is performed using the same optical bench as the one described in the work of Perin *et al.* [27], using a narrow linewidth external cavity diode laser at 420 nm which is continuously tunable over 2 GHz. The result is shown in Fig. 5. The measured curve is well-fitted using a theoretical model yielding an intrinsic Q-factor of $4.4 \times 10^7$. This high Q-factor of the cavity is beneficial for low threshold lasing.

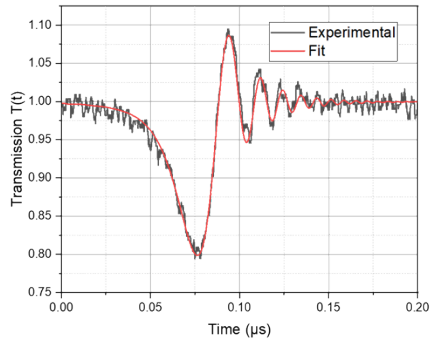

Fig. 5. Cavity ringdown measurement for the $Dy^{3+}$-doped glass microsphere, measured transmission (*black curve*) and fit (*red curve*). The model gives an intrinsic photon lifetime $\tau_0$ of 9.8 ns, and an extrinsic photon lifetime $\tau_e$ of 118.3 ns. The theoretical model and detailed fitting procedure can be found in Dumeige *et al.* [32].

## 3. Spectroscopic study

The µ-Raman and µ-luminescence spectroscopies are involved in characterizing the glass microspheres, see Fig. 6. Three different areas are considered (focusing on the equatorial plane): i) the center, ii) the junction with the fiber, and iii) the surface. The µ-Raman spectra, Fig. 6(a), are close for all three areas being in line with those for the fiber. It confirms the uniformity of the host glass composition in the microsphere and its minimal variation with respect to the fiber. The µ-luminescence spectra, Fig. 6(b), are also very similar to those for the fiber core. The luminescence intensity is slightly reduced close to the surface and increases significantly in the junction. The latter could be assigned to the accumulation of $Dy^{3+}$ ions in this defective area during the formation of the microsphere and suggests a possibly lower $Dy^{3+}$ doping level in the latter as compared to the fiber.

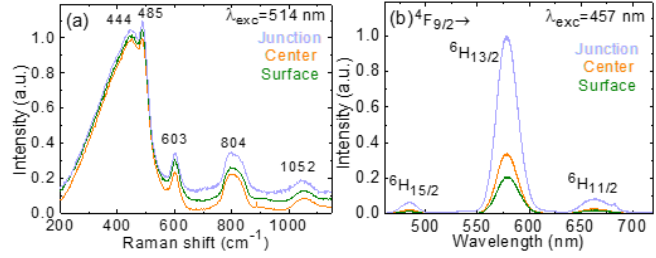

Fig. 6. Spectroscopy of the $Dy^{3+}$-doped glass microsphere: (a) normalized μ-Raman spectra, $\lambda_{exc}$ = 514 nm; (b) μ-luminescence spectra, $\lambda_{exc}$ = 457 nm (measured from the microsphere center, its junction with the fiber and the surface).

To characterize the homogeneity of $Dy^{3+}$ doping distribution in the microsphere, μ-luminescence mapping is performed (the spatial resolution is 0.5 μm). The yellow emission at 578 nm (the $^4F_{9/2} \rightarrow {}^6H_{13/2}$ transition) is monitored, and the emission peak intensity, peak position, and peak width are plotted as depicted in Fig. 7(a-c). The map of luminescence intensity highlights the high homogeneity of $Dy^{3+}$ distribution with only a slight decrease close to the surface and the junction (from the microsphere side). It also confirms the accumulation of $Dy^{3+}$ in the glass ripples in the junction. Near the surface, the yellow emission band of $Dy^{3+}$ experiences a red shift and a noticeable broadening. This could be attributed to the effect of the surface stress in the microsphere and (partially) to a slight variation in the glass composition.

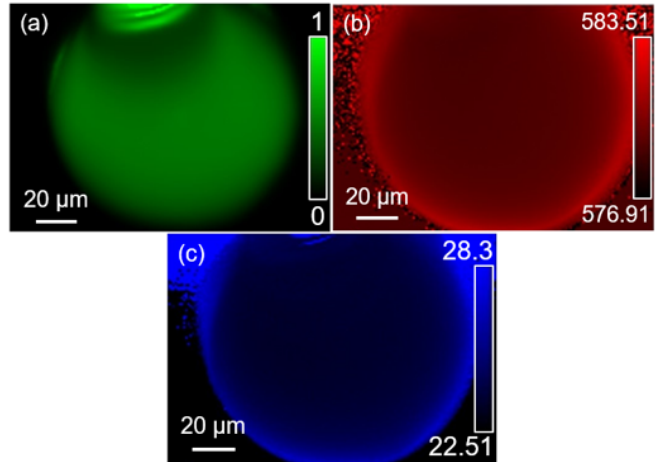

Fig. 7. μ-luminescence mapping of the $Dy^{3+}$-doped glass microsphere monitoring the yellow emission at ~578 nm: (a) peak intensity; (b) peak position and (c) peak width, $\lambda_{exc}$ = 457 nm.

The lifetime of the $^4F_{9/2}$ manifold of $Dy^{3+}$ ions is sensitive to the doping concentration as the luminescence originating from this level is subject to self-quenching via several phonon-assisted cross-relaxation processes. It could also be affected by possible energy migration to defects and impurities, depending on the glass host composition. Thus, the luminescence lifetime measurements could serve as a sensitive tool to compare the compositional variation between the glass and the microsphere.

The scheme of the setup used for the luminescence lifetime measurements is depicted in Fig. 8. The pump source is a blue 454-nm GaN laser diode (*Oxxius* Multimode Laser) operating around 454 nm (laser linewidth: 1.25 nm) addressing the $^6H_{15/2} \rightarrow {}^4I_{15/2}$ absorption band of $Dy^{3+}$ ions. Its output is modulated using a mechanical chopper (frequency: 100 Hz). The pump radiation is coupled into the 0.5-m long fiber using a spherical lens (focal length: $f$ = 40 mm) and the luminescence at the output fiber end-facet is collected using another lens ($f$ = 30 mm). The residual pump is filtered out using a long-pass filter (*Thorlabs*, FEL550). The luminescence is detected using a Si photodiode (response time: 16.5 µs) and an oscilloscope (*Tektronix MDO41043*). For the glass microsphere, a similar excitation scheme is used to illuminate the entire microresonator, and the luminescence is collected at a right angle using a ×40 microscope objective and detected as explained above. The excitation power is set to 70 mW.

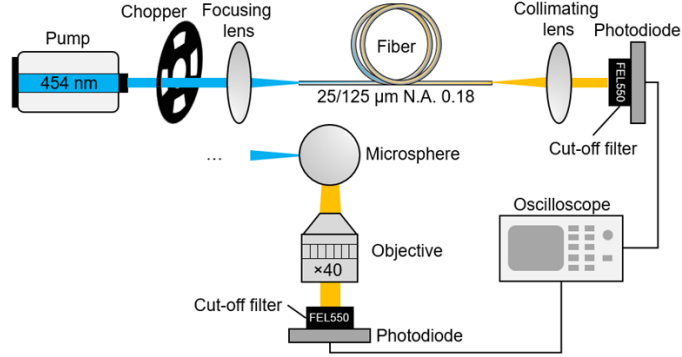

Fig. 8. Layout of the experimental setup for the luminescence lifetime measurements from the $Dy^{3+}$-doped fiber and glass microsphere.

The luminescence decay curves for both the fiber and the glass microsphere are shown in Fig. 9 plotted in a semi-log scale. The curves are well fitted using a single-exponential law yielding the luminescence lifetimes $\tau_{lum}$ of 500.8 ± 0.4 µs and 514.7 ± 1.1 µs, respectively. The $\tau_{lum}$ value for the fiber is in line with the previous report (~0.5 ms [30]). For the microsphere, we measure a slightly longer lifetime which could be assigned to both a reduction of the average doping level (*i.e.*, weaker self-quenching of luminescence) and a slight change in the host glass composition (due to mixing of the core and cladding fiber regions). The luminescence decay measurement is also performed focusing on the signal collected from the junction of the microresonator and the fiber strand giving a shorter value of $\tau_{lum}$ = 453.7 ± 1.3 µs. This supports the previous conclusion about the accumulation of dopant ions in this area.

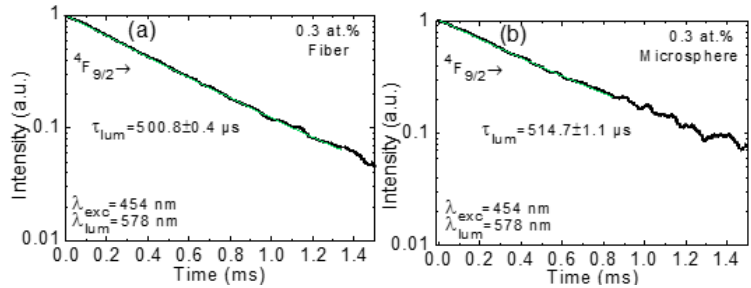

Fig. 9. Luminescence decay curves for $Dy^{3+}$ ions in (a) the pre-treated fiber and (b) glass microsphere (*r* = 60 µm), *circles* – experimental data, *lines* – their exponential fits. $\lambda_{exc}$ = 454 nm.

## 4. WGM microresonator

The high Q-factor of the spherical microresonator allows for a strong light confinement. The WGMs propagate by TIR along the surface of the microresonator, consequently, coupling light inside the cavity is a challenge. One efficient way to couple light is to make use of the evanescent wave by overlapping the evanescent tail of the external pump field with the WGM tail of the resonator. This can be achieved in different ways by utilizing either a prism [33], an angle polished fiber [34], a fiber full taper [35], or a fiber half taper [36].

In the present work, a fiber half taper is used to couple the pump light to the microresonator. A single-mode fiber (*Thorlabs*, 405-XP) served as a precursor for the fabrication of half-tapers. They were produced using a heat-and-pull method with a fiber fusion splicing apparatus (*Ericson*, FSU 925) resulting in taper diameters less than 1 μm. The tail of the evanescent field generated in the sub-micron part of the tapered fiber could be coupled to the microresonator by bringing it in proximity to the resonator. By an appropriate pump coupling, the fluorescence could also be made in resonance with the cavity modes. In this work, two separate half-tapers were used for pumping and collecting back the fluorescence /stimulated-emission signal.

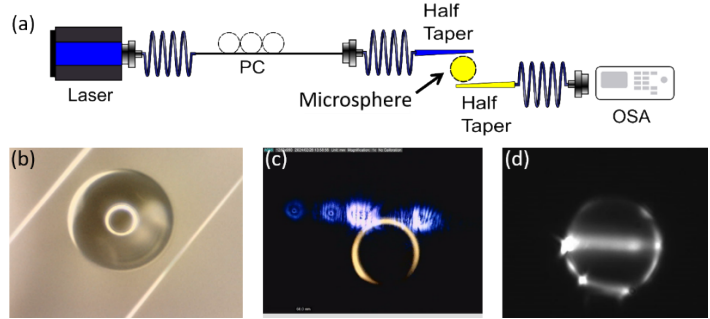

Fig. 10. (a) Optical bench used for the characterization of the WGM microresonators: PC - Polarization Controller; OSA - Optical Spectrum Analyzer; (b) a microsphere with a radius $r$ of 67 μm along with two half tapers placed inside an isolated cabin; (c) yellow fluorescence seen from the microsphere for 70 mW of pump power (top view); (d) WGMs on the microresonator observed with a long-pass filter (side view).

The experimental setup for the characterization of WGMs from $Dy^{3+}$-doped glass microspheres is depicted in Fig. 10(a). The 454 nm diode laser (*Oxxius*) is used as the pump source. Its emission is injected into the fiber half-taper with an inline polarization controller. The latter helped to select different cavity modes by varying the polarization of the injected modes. The two fiber half tapers and the microresonator are placed inside a cabin to isolate them from environmental perturbations. The precise positioning of fiber tapers is enabled via piezoelectric controllers. The output from the fiber taper used for collecting the WGM signal is directly connected to an optical spectrum analyzer (OSA, *Yokogawa*, AQ6373B). Inside the cabin, two cameras are placed to observe the microsphere, and both half tapers from top and side projections, see Fig. 10(b). This also helps in achieving the precise coupling between the microresonator and half tapers.

As a first step to effectively couple light inside the cavity, it is observed using a camera with a long-pass filter filtering out the blue pump light. This allows for observing the WGM fluorescence inside the microresonator, thus confirming effective coupling. Once the fluorescence is observed, cf. Fig. 10(c, d), the WGM fluorescence is collected by using another half taper and detected with the OSA.

Figure 11 depicts the fluorescence spectra in the yellow spectral range collected from the glass microresonator, measured with a resolution of 0.1 nm for various pump powers, i.e., 40 mW (clearly showing the filtered fluorescence by WGM resonant modes with a free spectral range (FSR) of 0.5 nm) and 70 mW (showing the onset of lasing). This supports the conclusion that the observed modes are WGMs supported by the microresonator. The cavity FSR was also calculated using the formula FSR = $c/(2\pi Nr)$, where $c$ is the speed of light in vacuum, $r$ is the radius of the spherical microresonator and $N$ is the group index (assumed to be 1.47), yielding 0.55 nm, which aligns well with the FSR observed in the spectrum. Increasing the pump power from 40 mW to 70 mW resulted in the enhancement of the WGM signal collected and the observation of the onset of lasing.

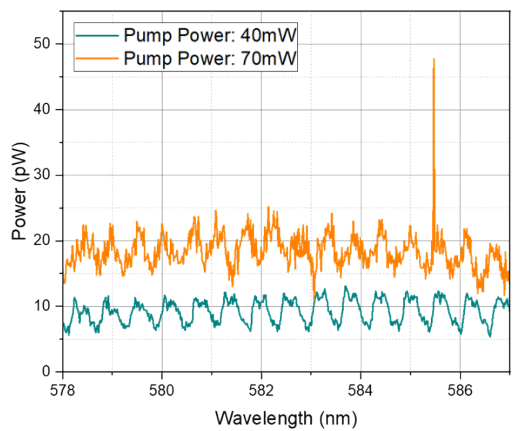

Fig. 11. Whispering gallery mode spectra from a $Dy^{3+}$-doped glass microresonator ($r = 67$ µm): the pump power is (*green*) 40 mW and (*orange*) 70 mW (onset of stimulated-emission). Spectral resolution: 0.1 nm, free spectral range (FSR) of the glass microcavity: 0.5 nm. The spectra are measured over 20 and 15 accumulations, respectively.

## 5. Conclusions

To conclude, we report on the first laser-active whispering gallery mode microresonator based on $Dy^{3+}$ ions as a potential source of narrow-linewidth emission in the yellow which is a challenging but highly demanded spectral range. We also present a proof-of-concept of a direct pumping scheme for visible microlasers, benefiting from the technology of blue GaN diode lasers. Based on the µ-Raman and µ-luminescence spectroscopy, as well as confocal laser microscopy, we conclude about the high homogeneity of the host composition, as well as the distribution of the dopant $Dy^{3+}$ ions across the microsphere which was derived from a partially stripped aluminosilicate glass fiber. Coupled with a high Q-factor of the derived microcavity, the utilized approach based on fiber fusion splicing appears as a viable solution for producing visible-emitting microlasers. This approach can be consequently extended to other rare-earth dopants providing direct emission in the visible under direct blue pumping such as $Pr^{3+}$, $Sm^{3+}$, or $Eu^{3+}$.

**Funding.** Indo-French Centre for the Promotion of Advanced Research (70T12-2, MASSALAQ), I-DEMO PIA4 Bpifrance - Région Bretagne – Lannion Tregor Communauté (40898814/1, QoQeliQo), Lannion Tregor Communauté – Région Bretagne (ARED) (ELVIS Project), Région Normandie, France (Contrat de plan État-Région (CPER)); Agence Nationale de la Recherche (ANR), France (ANR-22-CE08-0025-01, NOVELA).

**Disclosures.** The authors declare no conflicts of interest.

**Acknowledgment.** The authors thank Exail and Oxxius companies for providing the fiber samples and laser diode.

**Data availability.** Data underlying the results presented in this paper are not publicly available at this time but may be obtained from the authors upon reasonable request.

# References

1. T.-C. Wu, Y.-C. Chi, H.-Y. Wang, *et al.*, "Blue Laser Diode Enables Underwater Communication at 12.4 Gbps," Sci Rep **7**, 40480 (2017).
2. C. Toninelli, I. Gerhardt, A. S. Clark, *et al.*, "Single organic molecules for photonic quantum technologies," Nat. Mater. **20**, 1615–1628 (2021).
3. A. D. Ludlow, M. M. Boyd, J. Ye, *et al.*, "Optical atomic clocks," Rev. Mod. Phys. **87**, 637–701 (2015).
4. R. J. Niffenegger, J. Stuart, C. Sorace-Agaskar, *et al.*, "Integrated multi-wavelength control of an ion qubit," Nature **586**, 538–542 (2020).
5. S. M. Borizov and O. S. Wolfbeis, "Optical Biosensors," Chemical Reviews, **108**, 2 (2008)
6. K. Aikawa, J. Kobayashi, K. Oasa, *et al.*, "Narrow-linewidth light source for a coherent Raman transfer of ultracold molecules," Opt. Express, OE **19**, 14479–14486 (2011).
7. A. Chiasera, Y. Dumeige, P. Féron, *et al.*, "Spherical whispering-gallery-mode microresonators," Laser & Photonics Reviews **4**, 457–482 (2010).
8. T. J. Kippenberg, S. M. Spillane, and K. J. Vahala, "Demonstration of ultra-high-Q small mode volume toroid microcavities on a chip," Applied Physics Letters **85**, 6113–6115 (2004).
9. S. Zhu, L. Shi, L. Ren, *et al.*, "Controllable Kerr and Raman-Kerr frequency combs in functionalized microsphere resonators," Nanophotonics **8**, 2321–2329 (2019).
10. A. Rasoloniaina, V. Huet, M. Thual, *et al.*, "Analysis of third-order nonlinearity effects in very high-Q WGM resonator cavity ringdown spectroscopy," J. Opt. Soc. Am. B **32**, 370 (2015).
11. Q. Kuang, C. Xie, M. Wang, *et al.*, "Controllable Brillouin laser and Brillouin-Kerr microcombs," Opt. Express, OE **32**, 46698–46711 (2024).
12. B. Jiang, S. Zhu, L. Ren, *et al.*, "Simultaneous ultraviolet, visible, and near-infrared continuous-wave lasing in a rare-earth-doped microcavity," Adv. Photon. **4**, (2022).
13. J. Yu, X. Wang, W. Li, *et al.*, "An Experimental and Theoretical Investigation of a 2 μm Wavelength Low-Threshold Microsphere Laser," J. Lightwave Technol., **38**, 1880–1886 (2020).
14. V. Sandoghdar, F. Treussart, J. Hare, *et al.*, "Very low threshold whispering-gallery-mode microsphere laser," Phys. Rev. A **54**, R1777–R1780 (1996).
15. F. Lissillour, D. Messager, G. Stéphan, *et al.*, "Whispering-gallery-mode laser at 1.56 μm excited by a fiber taper," Opt. Lett., OL **26**, 1051–1053 (2001).
16. M. Cai, O. Painter, K. J. Vahala, *et al.*, "Fiber-coupled microsphere laser," Opt. Lett. **25**, 1430 (2000).
17. L. Mescia, P. Bia, O. Losito, *et al.*, "Design of Mid-IR $Er^{3+}$-Doped Microsphere Laser," IEEE Photonics Journal **5**, 1501308–1501308 (2013).
18. B. Behzadi, R. K. Jain, and M. Hossein-Zadeh, "Spectral and Modal Properties of a Mid-IR Spherical Microlaser," IEEE Journal of Quantum Electronics **53**, 1–9 (2017).
19. H. Fujiwara and K. Sasaki, "Upconversion lasing of a thulium-ion-doped fluorozirconate glass microsphere," Journal of Applied Physics **86**, 2385–2388 (1999).
20. W. von Klitzing, E. Jahier, R. Long, *et al.*, "Very low threshold green lasing in microspheres by up-conversion of IR photons," J. Opt. B: Quantum Semiclass. Opt. **2**, 204–206 (2000).
21. X. Wang, Y. Yu, S. Wang, *et al.*, "Single mode green lasing and multicolor luminescent emission from an $Er^{3+}$-$Yb^{3+}$ co-doped compound fluorosilicate glass microsphere resonator," OSA Continuum **1**, 261 (2018).
22. C. Kränkel, D.-T. Marzahl, F. Moglia, *et al.*, "Out of the blue: semiconductor laser pumped visible rare-earth doped lasers," Laser & Photonics Reviews **10**, 548–568 (2016).
23. Y. Fujimoto, J. Nakanishi, T. Yamada, *et al.*, "Visible fiber lasers excited by GaN laser diodes," Progress in Quantum Electronics **37**, 185–214 (2013).
24. G. Nunzi Conti, A. Chiasera, L. Ghisa, *et al.*, "Spectroscopic and lasing properties of $Er^{3+}$-doped glass microspheres," Journal of Non-Crystalline Solids **352**, 2360–2363 (2006).
25. L. Yang and K. J. Vahala, "Gain functionalization of silica microresonators," Opt. Lett., OL **28**, 592–594 (2003).
26. Z. Chen, X. Tu, J. Zhao, *et al.*, "An Erbium-Doped Fiber Whispering-Gallery-Mode Microcavity Laser," IEEE Photonics Technology Letters **31**, 1650–1653 (2019).
27. G. Perin, Y. Dumeige, P. Féron, *et al.*, "High-Q Whispering-Gallery-Modes Microresonators in the Near-Ultraviolet Spectral Range," J. Lightwave Technol., **42**, 15, 5214-5222 (2024).
28. T. Okazaki, C. Otsuka, E. H. Sekiya, *et al.*, "Diode pumped visible $Dy^{3+}$-doped silica fiber laser: Ge-co-doping effects on lasing efficiency and photodarkening," Appl. Phys. Express **15**, 012002 (2022).
29. J. Demaimay, E. Kifle, P. Loiko, *et al.*, "Efficient yellow Dy:ZBLAN fiber laser with high-brightness diode-pumping at 450 nm," Opt. Lett. **49**, 4174 (2024).
30. M. R. Majewski and S. D. Jackson, "Diode pumped silicate fiber for yellow laser emission," OSA Continuum, OSAC **4**, 2845–2851 (2021).
31. P. F. Mcmillan, B. T. Poe, and B. Reynard, "A study of $SiO_2$ glass and supercooled liquid to 1950 K via high-temperature Raman spectroscopy," Geochimica et Cosmochimica Acta, **58**, 17, 3683-3664 (1994).
32. Y. Dumeige, S. Trebaol, L. Ghişa, *et al.*, "Determination of coupling regime of high-Q resonators and optical gain of highly selective amplifiers," J. Opt. Soc. Am. B **25**, 2073 (2008).
33. M. L. Gorodetsky and V. S. Ilchenko, "Optical microsphere resonators: optimal coupling to high-Q whispering-gallery modes," J. Opt. Soc. Am. B **16**, 147 (1999).

34. V. S. Ilchenko, X. S. Yao, and L. Maleki, "Pigtailing the high-Q microsphere cavity: a simple fiber coupler for optical whispering-gallery modes," Opt. Lett. **24**, 723 (1999).
35. J. C. Knight, G. Cheung, F. Jacques, *et al.*, "Phase-matched excitation of whispering-gallery-mode resonances by a fiber taper," Opt. Lett. **22**, 1129 (1997).
36. A. Rasoloniaina, V. Huet, T. K. N. Nguyên, *et al.*, "Controlling the coupling properties of active ultrahigh-Q WGM microcavities from undercoupling to selective amplification," Sci Rep **4**, 4023 (2014).